\documentclass[a4paper,11pt]{article}
\usepackage{jheppub} 
\usepackage{tikz-cd}

\title{Chern Character for Discrete Spectrum Partition Function}

\author[1]{Shunrui Li} 
\author[2]{and Yang Liu}

\affiliation[1]{Department of mathematics, Free University of Berlin, Berlin 14195, Germany}
\affiliation[2]{Department of Physics, Tsinghua University, Beijing 100084, China}

\emailAdd{lis02@zedat.fu-berlin.de}
\emailAdd{liu-yang\_1990@mail.tsinghua.edu.cn}

\abstract{We establish a rigorous geometric correspondence between thermal partition functions of discrete-spectrum quantum systems with bounded ground energy and the Chern character of “virtual physical sheaf” over spacetime. By interpreting Hamiltonian dynamics as a $U(1)$-equivariant flow on the quantum phase space $\mathbb{CP}^n$ and pushforward to spacetime, we show that the finite-temperature partition function emerges as the integral of the Chern character of “virtual physical sheaf” over spacetime. The construction extends naturally to infinite dimensions through trace class guaranteed by Weyl’s aspmtotic law. Using the Grothendieck–Riemann–Roch formalism, we prove pushforward invariance of the Chern character under thermal compactification on arbitrary manifolds, providing a topological foundation for thermal traces in quantum field theory. This framework unifies spectral theory with characteristic class theory, offering a geometric interpretation of partition functions based on operator-algebraic approach.}

\begin{document}
\maketitle

\section{Introduction}
Systems with discrete energy levels and bounded ground state energy are foundational across all branches of quantum physics, but their thermal partition functions have long been treated as purely analytical quantities derived solely from spectral properties, with no connection to topological invariants of vector bundles or sheaves. In this work, we introduce the “virtual physical sheaf", a formal topological construct whose underlying fibers intrinsically encode the space of quantum states, and rigorously establish that the partition function of any such system is exactly the integral of the Chern character of this sheaf over spacetime.

To implement this topological mapping rigorously, we employ a spectral filtration approach. By truncating the system to its first $n+1$ eigenstates, we identify the finite-dimensional projective phase space with $\mathbb{CP}^n$. Here, the truncated partition function emerges precisely as the integral of an equivariant Chern character, localized via the Atiyah-Bott theorem. Protected by the strict trace-class property of the Euclidean evolution operator $e^{-\beta H}$ (governed by Weyl's asymptotic law), this geometric correspondence safely and absolutely converges in the infinite-dimensional limit. Furthermore, using the Grothendieck-Riemann-Roch formalism, we demonstrate the pushforward invariance of the Chern character under thermal compactification, revealing that the standard Matsubara frequency summation inherently corresponds to a geometric pushforward operation. The paper develops this core correspondence in Section 2 and discusses its implications in Section 3. Appendices provide technical details on trace-class operators and equivariant characteristic class identities.

\section{Partition function and the Chern character}
In this section, we consider quantum systems whose Hamiltonian \(H\) has a purely discrete spectrum \(\{E_n\}_{n=0}^{\infty}\) that is bounded from below, i.e. \(E_0 > -\infty\). For such systems, and the partition function at inverse temperature \(\beta\) is a finite, well-defined scalar:
\begin{equation} \label{def Z}
    Z(\beta) = \mathrm{Tr}_{\mathcal{H}}\big[e^{-\beta H}\big] = \sum_{n} e^{-\beta E_n},
\end{equation}
where the sum runs over the discrete eigenvalues of the Hamiltonian. The exponential \(e^{-\beta H}\) is the unnormalized density matrix of the canonical ensemble.

The Chern character of a complex vector bundle \(E \to M\) with connection \(\nabla\) is the differential form
\begin{equation} \label{def ch}
    \mathrm{ch}(E) = \mathrm{Tr}\!\left[\exp\!\left(\frac{i}{2\pi}F\right)\right],
\end{equation}
where \(F = \nabla^2\) is the curvature two-form and the trace is taken over the fibre indices.

Formally, \eqref{def ch} and \eqref{def Z} share the same algebraic skeleton: both are a trace of an exponential. In the Chern character, the exponential contains the curvature \(F\), while in the partition function, the exponential contains the Hamiltonian \(H\). The structural analogy
\begin{equation} \label{structural analogy}
    \mathrm{Tr}\big[e^{-\beta H}\big] \;\longleftrightarrow\; \mathrm{Tr}\big[e^{F}\big]
\end{equation}
is algebraically striking. 
This inspires us to explore the connection between the Chern character and the partition function, which is the problem our paper aims to address.

Next, we formalize the partition function of systems with discrete energy levels and bounded ground state energy as the geometric evaluation of a “virtual physical sheaf" $\mathcal{S}_M$ over spacetime $M$. We simply define $\mathcal{S}_M$ as a formal topological construct that intrinsically encodes the quantum state space. To see how the Hamiltonian dynamics is geometrically captured, consider a general quantum state $|\psi\rangle$. Utilizing the complete orthonormal basis of energy eigenstates $\{|\psi_i\rangle\}$, any state admits a unique expansion:
\begin{equation} \label{expansion}
    \psi(x)=\sum_n c_i \psi_i(x),
\end{equation}
where \( c_i = \langle \psi_i(x), \psi(x) \rangle_{L^2} \) are the expansion coefficients, and $\psi(x)$ is normalized wave function. 

Furthermore, this expansion is expected to converge in the \( L^2 \) sense, meaning that:
\begin{equation} \label{L2 convergent}
    \lim_{n \to \infty} \left\| \psi(x) - \sum_{i=0}^{n} c_i \psi_i(x) \right\|_{L^2} = 0.
\end{equation}
We begin by considering the Hamiltonian operator $H$ as a linear map acting on the sections of the virtual physical sheaf:
\begin{equation} \label{section H A1}
    H: \Gamma(M, \mathcal{S}_M) \rightarrow \Gamma(M, \mathcal{S}_M).
\end{equation}
We regard the Hamiltonian as a mapping on the quantum state space. Geometrically, this means $H$ acts directly on the sections of $\mathcal{S}_M$ over the spacetime manifold $M$. That is, for any section $\psi(x,t) \in \Gamma(M, \mathcal{S}_M)$ representing a quantum state, the Hamiltonian maps it via:
\begin{equation}\label{section H A2}
   \psi(x,t) \mapsto H\psi(x,t).
\end{equation}
Our next step is to clarify the relation between “virtual physical sheaf" $\mathcal{S}_M$ and the discrete spectral structure of the Hamiltonian.
\subsection{Finite dimensional case}
To truncate the spectrum at the first $n+1$ levels, consider all wave functions of the form
$\displaystyle \psi(x)=\sum_{i=0}^n a_i\,\psi_i(x)$
with complex coefficients $a_i$. 
 By normalizing the wave functions and quotienting out the unobservable global phase, the $(n+1)$-dimensional complex Hilbert space naturally projects onto the complex projective space $\mathbb{CP}^n$. Formally,
the complex projective space \(\mathbb{CP}^n\) can be defined as:
\begin{equation} \label{CPn}
\mathbb{CP}^n = \frac{\mathbb{C}^{n+1} \setminus \{0\}}{\mathbb{C}^*},
\end{equation}
where \(\mathbb{C}^{n+1}\) is a complex vector space of \((n+1)\)-dimension, and \(\mathbb{C}^*\) denotes the multiplicative group of non-zero complex numbers. Each point \([z] \in \mathbb{CP}^n\) corresponds to an equivalence class of nonzero vectors under scalar multiplication, known as a homogeneous coordinate \cite{nakahara2018geometry}.

In the framework of geometric quantization, the classical phase space is endowed with a symplectic structure \cite{brody2001geometric, ashtekar1999geometrical}.
In this framework, the true physical state space of the finite-level quantum system is exactly the K\"{a}hler manifold $\mathbb{CP}^n$. Every pure quantum state corresponds simply to a generic point on this symplectic manifold, and the standard Fubini-Study form $\omega_{FS}$ serves as the symplectic structure.

Over this projective phase space, we introduce the prequantum line bundle $\mathcal{O}(1)$. As formulated in the geometric quantization framework and utilized by Atiyah and Bott \cite{atiyah1984moment}, this line bundle is introduced specifically because its equivariant Chern character canonically encodes the Hamiltonian moment map $\mu = \beta H$. 

The thermodynamic evolution of the system is captured by a continuous $U(1)$ geometric flow on this projective phase space, generated by the Hamiltonian operator $H$. Under the evolution operator $e^{-\beta H}$, a generic state traverses an orbit, whereas an energy eigenstate merely acquires a scalar phase $e^{-\beta E_i}$. Because global phases are identified as single points in projective space, these eigenstates experience no geometric displacement. Consequently, the discrete energy eigenstates correspond identically to the fixed points $p_i$ of this $U(1)$ flow \cite{ashtekar1999geometrical, brody2001geometric}. This natural Hamiltonian action allows us to directly invoke the equivariant machinery of Atiyah and Bott \cite{atiyah1984moment}: it uniquely defines a moment map $\mu$ that physically represents the generating observable, evaluating in our thermodynamic setting exactly to the scaled quantum Hamiltonian, $\mu = \beta H$. To construct a globally conserved topological invariant under this thermal evolution, the ordinary symplectic form is promoted via the Cartan model to an equivariantly closed form. This effectively serves as the equivariant curvature $F_{eq}$ of the prequantum line bundle $\mathcal{O}(1)$, seamlessly coupling the geometric phase space with the spectral energy data \cite{BNGE VM 1992}:
\begin{equation} \label{equivariant_form}
\omega_{eq} = \omega_{FS} - \beta H u,
\end{equation}
where $u$ parameterizes the circumference of the $S^1$ action \footnote{From the perspective of algebraic topology, one might question how a 0-form (the scalar function $H$) can be subtracted from a 2-form ($\omega_{FS}$). In the Cartan model of equivariant cohomology \cite{tu2020introductory}, the parameter $u$ (playing the role of the generator of the $U(1)$ equivariant polynomial ring) is formally assigned a topological degree of 2. This grading precisely ensures that the equivariant symplectic form $\omega_{eq} = \omega_{FS} - \beta H u$ is homogeneously of degree 2, resolving the apparent geometric mismatch.}. Next, based on the definition, we introduce equivariant curvature \cite{BNGE VM 1992}: 
\begin{equation} \label{equivariant_curvature}
F_{eq} = -\omega_{eq}.
\end{equation}
Then we have the equivariant Chern character \cite{BNGE VM 1992}:
\begin{equation} \label{equivariant_chern_character}
\mathrm{ch}_{eq} =\operatorname{Tr}[\operatorname{exp}(-F_{eq})]= \operatorname{Tr}[\operatorname{exp}(\omega_{eq})] = \operatorname{Tr}[\operatorname{exp}(\omega_{FS} - \beta H u)].
\end{equation}
Then we apply the Atiyah-Bott localization theorem (for the derivation of the equivariant Euler class, refer to Appendix B) \cite{atiyah1984moment, tu2020introductory, BNGE VM 1992, DH 1982}:
\begin{equation}
    \int_{\mathbb{CP}^n} e^{\omega_{FS}-\beta Hu}=\sum_{i=0}^n \frac{e^{-\beta E_i}}{\prod_{j \neq i}(-\beta(E_j-E_i))}
\end{equation}

To rigorously map the abstract geometric structures from the quantum phase space $\mathbb{CP}^n$ down to the physical spacetime manifold $M$, we must recognize that the process of geometric quantization over families is mathematically governed by the K-theoretic pushforward. Let us consider the projection map $\pi: M \times \mathbb{CP}^n \to M$. We construct the “virtual physical sheaf" $\mathcal{S}_M$ on the spacetime $M$ via the pushforward $\mathcal{S}_M = \pi_! \mathcal{O}(1)$.
However, the Chern character is famously not invariance under pushforwards. 

To correctly map the topological invariants between these different manifolds, we are strictly bound by the Grothendieck-Riemann-Roch (GRR) theorem for the projection $\pi$. Applying the GRR theorem to the total tangent bundles yields:
\begin{equation} \label{GRR general}
\mathrm{ch}(\pi_! \mathcal{O}(1)) \wedge \mathrm{Td}(T_M) = \pi_* \left( \mathrm{ch}_{eq}(\mathcal{O}(1)) \wedge \mathrm{Td}_{eq}(T_{M \times \mathbb{CP}^n})\right).
\end{equation}
To see how this powerful theorem governs our physical framework, we can evaluate the right-hand side step by step. First, for the trivial fibration $\pi: M \times \mathbb{CP}^n \to M$, the tangent bundle of the total space naturally splits into the pullback of the base and the tangent space of the fiber:
\begin{equation}
T_{M \times \mathbb{CP}^n} = \pi^*T_M \oplus T_{\mathbb{CP}^n}.
\end{equation}
Since the $U(1)$ thermal action operates exclusively on the phase space $\mathbb{CP}^n$ and trivially on the spacetime manifold $M$, the equivariant Todd class respects this splitting via its multiplicative property:
\begin{equation}
\mathrm{Td}_{eq}(T{M \times \mathbb{CP}^n}) = \pi^* \mathrm{Td}(T_M) \wedge \mathrm{Td}_{eq}(T{\mathbb{CP}^n}).
\end{equation}
Substituting this factorization back into the right-hand side of \eqref{GRR general}, we apply the fundamental projection formula for fiber integration, $\pi_*(\alpha \wedge \pi^*\beta) = \pi_*(\alpha) \wedge \beta$. This allows the macroscopic Todd class to smoothly factor out of the pushforward operation:
\begin{equation}
\text{RHS} = \pi_* \left( \mathrm{ch}_{eq}(\mathcal{O}(1)) \wedge \mathrm{Td}_{eq}(T_{\mathbb{CP}^n}) \right) \wedge \mathrm{Td}(T_M).
\end{equation}
Comparing this directly with the left-hand side of \eqref{GRR general}, the term $\mathrm{Td}(T_M)$ precisely cancels out from both sides. By geometrically identifying the pushforward $\pi_! \mathcal{O}(1)$ with the virtual physical sheaf $\mathcal{S}_M$ over spacetime, the theorem elegantly collapses into the exact topological mapping:
\begin{equation} \label{GRR specific}
\mathrm{ch}(\mathcal{S}_M) = \pi_* \left( \mathrm{ch}_{eq}(\mathcal{O}(1)) \wedge \mathrm{Td}_{eq}(T_{\mathbb{CP}^n}) \right).
\end{equation}
Then we combine with \eqref{app_euler} and \eqref{app_todd} and use the Atiyah-Bott localization theorem again:
\begin{equation} \label{int}
\begin{split}
    \int_{\mathbb{CP}^n}(\mathrm{ch}_{eq}(\mathcal{O}(1)) \wedge \mathrm{Td}_{eq}(T_{\mathbb{CP}^n}))&=\sum_{i=0}^n\left(\frac{e^{-\beta E_i}}{\prod_{j \neq i}(-\beta(E_j-E_i))} \prod_{i\neq j}\frac{-\beta(E_j-E_i)}{1-e^{-\beta(E_j-E_i)}}\right)\\
    &=\sum_{i=0}^n \left(\frac{e^{-\beta E_i}}{\prod_{j \neq i}(1-e^{-\beta(E_j-E_i))}}\right) 
\end{split}
\end{equation}
According to the formula \eqref{Residue equation} proven in the appendix, this integral is the partition function. Then we combine \eqref{int} with \eqref{GRR specific} and obtain that:
\begin{equation}
    Z_n(M) = \int_M \mathrm{ch}(\mathcal{S}_M) = \int_{M \times \mathbb{CP}^n} \mathrm{ch}_{eq}(\mathcal{O}(1)) \wedge \mathrm{Td}_{eq}(T_{\mathbb{CP}^n}) = \sum_{i=0}^n e^{-\beta E_i}.
\end{equation}
\subsection{Infinite dimensional case}
In the limit $n\to \infty$, the formal correspondence yields an expression that converges to the partition function $Z(M)$, as guaranteed by the trace-class property of $\text{e}^{-\beta H}$ (for details regarding operator algebra, please refer to the appendix A). Based on equations \eqref{expansion}, we can introduce a cutoff at $n$ for \(|\phi\rangle\). 
\begin{equation}
    |\psi \rangle=\lim_{n \to \infty} \left(\sum_{i=0}^{n} c_i \psi_i(x) \right).
\end{equation}

For systems in thermal quantum field theory characterized by discrete energy levels and a lower bound on energy, this expression converges as \(n\to\infty\), yielding a well-defined infinite-dimensional limit.

In Section 2.1, we studied quantum states through a spectral filtration, where the system is truncated to the first $n+1$ eigenstates and mapped to the projective space $\mathbb{CP}^n$. This mapping induces a natural geometric identification of the truncated partition function $Z_n(M)$ with the integral of an equivariant Chern character of the line bundle $\mathcal{O}(1)$ over $\mathbb{CP}^n$. As $n \to \infty$, we extend this correspondence to the full spectral limit. By exploiting the trace-class property of the Euclidean evolution operator $e^{-\beta H}$ (which ensures the absolute convergence of the spectral sum), we demonstrate that the sequence of geometric integrals $Z_n(M)$ rigorously converges to the physical partition function $Z(M)$:
\begin{equation} \label{limitation}
\lim_{n \to \infty} \int_{\mathbb{CP}^n} \mathrm{ch}_{eq}(\mathcal{O}(1)) = \text{Tr}[e^{-\beta H}] = \sum_{i=0}^{\infty} e^{-\beta E_i}.
\end{equation}
We adopt the following convention: the symbol \(\operatorname{ch}(\mathcal{S}_M)\) denotes the Chern character of the virtual physical sheaf as a cohomology class, and the partition function \(Z(M)\) is chosen as a concrete representative of this class. That is,
\begin{equation} \label{partition function= chern character A}
   \int_M\operatorname{ch}(\mathcal{S}_M)=\operatorname{Tr}[\text{exp}(-\beta H)]=\sum_{i=0}^{\infty}e^{-\beta E_i}=Z(M).
\end{equation}
Here the equality \(\int_M\operatorname{ch}(\mathcal{S}_M) = Z(M)\) is to be understood in the sense that the scalar partition function \(Z(M)\) furnishes a scalar representative of the Chern character cohomology class associated with the “virtual physical sheaf" \(\mathcal{S}_M\). The choice of \(Z(M)\) as the scalar representative is justified by the trace-class property of \(e^{-\beta H}\) established in Appendix A: the trace is finite, basis-independent, and uniquely determined by the spectrum of the Hamiltonian, hence it provides a well-defined scalar that can be consistently chosen to represent the underlying cohomology class.

This convention is directly analogous to the standard practice in index theory, where the index of an operator, a scalar integer, is identified with the evaluation of a characteristic class on the fundamental cycle of the underlying manifold. A topological invariant that originates as a cohomology class is represented by a concrete scalar quantity obtained via a trace or integral. When \(\operatorname{ch}(\mathcal{S}_M)\) appears in the cohomological operations of Sections 2.3 and 2.4, which include pullbacks, pushforwards, and multiplication by the Todd class, it is the underlying characteristic form that is being manipulated, while the scalar representative $Z(M)$ emerges as its evaluation after the quantum trace. Conversely, when only the numerical value of the partition function is needed, the scalar representative \(Z(M)\) is used directly. The finite and absolutely convergent partition function therefore yields a well-defined cohomology class, of which \(Z(M)\) is the chosen representative.

Moreover, in the previous statement, there is no requirement for $E_i$ to correspond to a non-degenerate state. Thus, if the partition function takes the form:
\begin{equation} \label{non-degenerate state}
    Z(M)=\sum_i C_i e^{-\beta E_i},
\end{equation}
where $C_i$ denotes the degeneracy of the quantum state, then, according to the above derivation, this expression for the partition function corresponds to the Chern character.

More generally, if each energy level $E_n$ carries the degeneracy $C_n$, one simply replaces $e^{-\beta E_n}$ in \eqref{partition function= chern character A} by $C_ne^{-\beta E_n}$, leading to:
\begin{equation} \label{non-degenerate state and chern character A}
    Z(M)=\sum_n C_n e^{-\beta E_n}=\operatorname{Tr}e^{-\beta H}=\operatorname{ch}(\mathcal{S}_M).
\end{equation}
In this form, the partition function $Z(M)$ defined on $M$ is formally isomorphic to the scalar representative of Chern character $\mathrm{ch}(\mathcal{S}_M)$ of “virtual physical sheaf" $\mathcal{S}_{M}$ . 

\subsection{Thermal pushforward and the Euclidean Heisenberg evolution}

In finite-temperature quantum field theory, the temporal direction is compactified into a thermal circle \(S^1_\beta \simeq \mathbb{R}/(\beta\mathbb{Z})\) \cite{TM 1955}. While the preceding sections employed a general spacetime language. In the remainder of this paper, we restrict our attention to quantum systems defined on a flat Euclidean spacetime. Let \(\Sigma\) denote the spatial manifold on which the quantum system is defined at zero temperature. The corresponding zero-temperature Euclidean spacetime is \(\Sigma \times \mathbb{R}\), and at finite inverse temperature \(\beta\) the Euclidean time direction is compactified, yielding the finite-temperature Euclidean spacetime \(\Sigma \times S^1_\beta\). All geometric and topological constructions in this and the following section are understood in the Euclidean signature.

To examine the topological effect of introducing a thermal bath, we consider the commutative diagram:

\begin{equation} \label{commutative_diagram}
\begin{tikzcd}
\Sigma \arrow[r, "\iota"] & \Sigma \times \mathbb{R} \arrow[r, "\sigma"] \arrow[d, "\pi_0"] & \Sigma \times S^1_{\beta} \arrow[d, "\pi_{\beta}"] \\
                          & \mathbb{R} \arrow[r, "c"]                                       & S^1_{\beta}                                
\end{tikzcd}
\end{equation}
where:
\begin{itemize}
    \item \(\Sigma\): spatial manifold (Euclidean signature);
    \item \(\Sigma \times \mathbb{R}\): zero-temperature Euclidean spacetime;
    \item $S^1_\beta$: thermodynamic circle with circumference $\beta = 1/(kT)$;
    \item $\Sigma \times S^1_\beta$: thermodynamic extended spacetime;
    \item $\iota: \Sigma \hookrightarrow \Sigma \times \mathbb{R}$ is the constant section embedding at Euclidean time $\tau = 0$, i.e. $\iota(x) = (x,0)$;
   \item $\sigma: \Sigma \times \mathbb{R} \to \Sigma \times S^{1}_{\beta}$: thermal compactification map that identifies Euclidean time $\tau \in \mathbb{R}$ with $\tau+\beta$;
    \item $\pi_0: \Sigma \times \mathbb{R} \to \mathbb{R}$ is the projection onto the Euclidean time direction at zero temperature;
    \item $\pi_\beta: \Sigma \times S^1_\beta \rightarrow S^1_\beta$:  projection onto the thermal circle;
   \item $c: \mathbb{R} \to S^1_{\beta}$ is the thermal compactification map, identifying $\tau \in \mathbb{R}$ with $\tau + \beta$.
\end{itemize}

The thermal extension of the quantum state space is realized by the pushforward of the “virtual physical sheaf" on zero-temperature:
\begin{equation}
    \mathcal{S}_\beta := \sigma_*\mathcal{S}_0,
\end{equation}
where $\mathcal{S}_0$ is the “virtual physical sheaf” defined on the zero-temperature Euclidean spacetime $\Sigma \times \mathbb{R}$ and $\sigma_*$ is the pushforward from $\Sigma \times \mathbb{R}$ to $\Sigma \times S_{\beta}^1$. At zero temperature ($\beta \rightarrow \infty$), the thermal circle decompactifies into the flat real line $\mathbb{R}$, yielding the conventional quantum mechanical state space. 

When the system is coupled to a thermal bath, the pushforward \(\sigma_*\) lifts the zero-temperature sheaf \(\mathcal{S}_0\) to the finite-temperature “virtual physical sheaf" \(\mathcal{S}_\beta\) on \(\Sigma \times S^1_\beta\). This geometric operation is the exact algebraic counterpart of the Euclidean Heisenberg picture in thermal quantum field theory: a static local operator \(\mathcal{O}(x)\) defined on the spatial manifold \(\Sigma\) is extended along the imaginary time direction by the Heisenberg evolution driven by the Hamiltonian \(H\):
\begin{equation} \label{heisenberg_euclidean}
    \mathcal{O}(x, \tau) = e^{\tau H} \mathcal{O}(x) e^{-\tau H},
\end{equation}
which naturally expands these pushforward sections into the discrete Matsubara frequency basis 
\cite{TM 1955}:
\begin{equation} \label{matsubara_pushforward}
    \sigma_*\mathcal{O}(x) \longrightarrow \sum_{n=-\infty}^{\infty} \mathcal{O}_n(x) e^{i\omega_n \tau}, \quad \text{where} \quad \omega_n = \frac{2\pi n}{\beta}.
\end{equation}
This explicit decomposition demonstrates exactly how the pushforward inherently equips the purely spatial quantum states with the imaginary-time dynamics and Matsubara frequencies, without altering the underlying spatial topology.

We will demonstrate the invariance of the Chern character under this thermal pushforward in the next subsection.
\subsection{Grothendieck-Riemann-Roch theorem and the general thermal trace}

In Section 2.3, we established the relation \(\text{ch}(\sigma_*\mathcal{S}_0) = \sigma_*\text{ch}(\mathcal{S}_0)\) under the thermal pushforward mapping. While this identity beautifully aligns with the naturality of characteristic classes, its deepest physical and geometric significance is fully revealed through the Grothendieck-Riemann-Roch (GRR) theorem.

Let \(f: X \rightarrow Y\) be a proper morphism between compact complex manifolds, and \(\mathcal{F}\) be a coherent sheaf on \(X\). The GRR theorem connects the topological index with the analytic pushforward via the following cohomology equation:
\begin{equation} \label{GRR}
    f_*(\text{ch}(\mathcal{F}) \cdot \text{Td}(T_X)) = \text{ch}(Rf_*\mathcal{F}) \cdot \text{Td}(T_Y),
\end{equation}
where \(\text{Td}(T)\) is the Todd class of the tangent bundle, and \(Rf_*\) represents the derived direct image sheaf of \(\mathcal{F}\) \cite{DSF2021}. 

In our thermodynamic framework, we formally apply the GRR theorem directly to the cross-sectional pushforward mapping \(\sigma: \Sigma \times \mathbb{R} \rightarrow \Sigma \times S^1_\beta\) for our zero-temperature spatial sheaf \(\mathcal{S}_0\). Substituting these into the GRR formula \eqref{GRR}, we obtain:
\begin{equation} \label{M S GRR}
    \sigma_*(\text{ch}(\mathcal{S}_0) \cdot \text{Td}(T_{\Sigma \times \mathbb{R}})) = \text{ch}(R\sigma_*\mathcal{S}_0) \cdot \text{Td}(T_{\Sigma \times S^1_\beta}).
\end{equation}

To evaluate this for the thermal quantum field theory, we follow the standard paradigm established in the classic computations of the Atiyah-Singer index theorem \cite{atiyah2006elliptic, atiyah1963index, ABP1973}: a quantum system defined on Euclidean space is first compactified to a sufficiently large torus to obtain a compact manifold on which the index theorem applies directly, the topological invariant is computed, and the thermodynamic limit is then taken to recover the flat-space result. This procedure is rigorous and has been the foundation of index-theoretic computations in quantum field theory since the 1960s \cite{atiyah2006elliptic, atiyah1963index, ABP1973}.

Based on our previous settings, the spatial manifold \(\Sigma\) is flat Euclidean space. After compactification, its tangent bundle is trivial, and the Todd class is geometrically forced to be unity: \(\mathrm{Td}(T_{\Sigma}) = 1\). The zero-temperature Euclidean time direction \(\mathbb{R}\) is likewise flat and, being one-dimensional, has trivial Todd class: \(\mathrm{Td}(T_{\mathbb{R}}) = 1\). The thermal circle \(S^1_\beta\) is also flat and parallelizable, giving \(\mathrm{Td}(T_{S^1_\beta}) = 1\). By the multiplicativity of the Todd class for product manifolds, we have
\begin{equation} \label{todd}
   \mathrm{Td}(T_{\Sigma \times \mathbb{R}}) = \mathrm{Td}(T_\Sigma) \cdot \mathrm{Td}(T_{\mathbb{R}}) = 1,
\qquad
\mathrm{Td}(T_{\Sigma \times S^1_\beta}) = \mathrm{Td}(T_\Sigma) \cdot \mathrm{Td}(T_{S^1_\beta}) = 1.
\end{equation}

These equalities are not assumptions but geometric consequences of the flatness of the underlying manifolds. Substituting them into the GRR formula, the relation reduces precisely to the identity established in the previous section:
\begin{equation} \label{conclusion GRR}
    \sigma_*\text{ch}(\mathcal{S}_0) = \text{ch}(R\sigma_*\mathcal{S}_0).
\end{equation}
Thus, for the thermal quantum field theory, the GRR theorem guarantees that the Chern character of the “virtual physical sheaf” on zero-temperature is invariant under thermal pushforward. This provides the topological basis for the identification of the thermal partition function as the Chern character scalar representative developed in Section 2.2. 

\section{Conclusions and discussions}
The findings presented in this work provide a geometric reinterpretation of the partition function for quantum systems with discrete energy levels and bounded ground state energy. By constructing the virtual physical sheaf over spacetime and applying the Atiyah-Bott localization theorem, we establish that the thermal partition function coincides exactly with the integral of the Chern character of this sheaf. This correspondence is rigorously extended to infinite dimensions via the trace-class property of $e^{-\beta H}$ (guaranteed by Weyl’s asymptotic law).

Our results connect several formal structures in physics and mathematics:
\begin{itemize}
  \item The thermal partition function is identified as the scalar representative of the Chern character of the virtual physical sheaf $\mathcal{S}_M$ over spacetime $M$;
  \item The Grothendieck-Riemann-Roch (GRR) theorem governs the compatibility of this pushforward with the Todd class, reducing to the pushforward identity $\sigma_*\text{ch}(\mathcal{S}_0) = \text{ch}(R\sigma_*\mathcal{S}_0)$ in flat spacetime.
\end{itemize}

This framework does not replace the conventional operator-algebraic approach to partition functions; rather, it provides a complementary geometric language that reorganizes the same spectral data into a topological context. The methods developed in this paper are not limited to some specific systems. Any quantum systems with discrete energy levels and bounded ground state energy, including harmonic oscillators, quantum dots, and many-body systems with confining potentials etc., should admit a similar topological description. This is because the key mathematical properties required for our construction, discrete spectrum and trace-class Euclidean evolution operator, are shared by all such systems.

Looking forward, this perspective opens several promising avenues for theoretical exploration. First, extending the virtual physical sheaf framework to many-body systems is a natural next step. This could reveal new topological phenomena in collective quantum systems and provide a unified geometric language for describing finite-temperature many-body physics. Second, while we have focused on flat spacetime in this work, the GRR theorem naturally extends to curved spacetimes, where non-trivial Todd classes will modify the thermodynamic quantities in a topologically protected manner. Explicit calculations in non-trivial geometries could lead to new insights into quantum thermodynamics in gravitational backgrounds. Third, investigating the physical implications of the topological nature of the partition function is an important direction. In particular, it would be interesting to explore whether topological invariants can be used to classify phase transitions.

In summary, we have demonstrated that the partition function of discrete-spectrum systems admits a geometric realization as a Chern character integral. This bridges statistical mechanics and algebraic topology, offering a unified framework for analyzing thermal traces in quantum theory.

\appendix

\section{The trace class}

To establish a formal geometric isomorphism between the the Chern character and the quantum partition function, we must systematically address the topological obstruction inherent in generalized quantum fields. According to Kuiper's theorem, the unitary group $G(\mathcal{H})$ of an infinite-dimensional separable Hilbert space is contractible \cite{K1965}. Consequently, Consequently, any vector bundle constructed over the spacetime manifold with the monolithic, un-truncated Hilbert space $\mathcal{H}$ as its fiber is topologically trivial, because its structure group $G(\mathcal{H})$ is contractible. This forces all its high-level Chern classes ($c_1, c_2, \dots$) to be vanished. 

In order to extract meaningful topological invariants, we cannot treat the entire infinite-dimensional $\mathcal{H}$ as the singular geometric fiber of the physical sheaf; instead, we must spectrally filter the sheaf by restricting its fibers to topologically non-trivial, finite-dimensional components spanned by the eigenspaces.

For the class of quantum systems considered in this paper whose Hamiltonian $H$ has a purely discrete spectrum $\{E_n\}_{n=0}^\infty$ bounded from below, i.e. $E_0 > -\infty$, the physical justification for this filtration emerges naturally from the spectral properties of $H$. Crucially, for a physically well-behaved Hamiltonian acting as an elliptic operator, the discrete spectrum satisfies Weyl's asymptotic law. This mathematically dictates that the energy eigenvalues grow asymptotically as $E_n \sim C n^{2/d}$ for large $n$, where $d$ is the dimension of the manifold and $C > 0$ is a positive constant \cite{Weyl1911}.

Since the spectrum is discrete, bounded from below, and strictly governed by this asymptotic growth, the Euclidean evolution operator $e^{-\beta H}$ is a positive self-adjoint operator whose eigenvalues satisfy $0 < e^{-\beta E_n} \leq e^{-\beta E_0} < \infty$. Consequently, $e^{-\beta H}$ is not merely a bounded operator, but a compact operator, ensuring the continuous part of the spectrum is absent. Furthermore, because the high-energy contributions are exponentially suppressed by the thermal factor $\beta$, the sequence of eigenvalues scales as $e^{-\beta E_n} \sim \exp(-\beta C n^{2/d})$. This fractional-exponential decay drops off faster than any polynomial, thereby guaranteeing the absolute convergence of the infinite sum $\sum e^{-\beta E_i}$. Therefore, the operator $e^{-\beta H}$ safely and rigorously descends into the trace class \cite{CA1994, SB2005}.

This operator-theoretic classification is explicitly verified by evaluating its trace. By definition, a positive, self-adjoint operator belongs to the trace-class if the sum of its eigenvalues absolutely converges. For any system in the class defined above,:
\begin{equation}
    \mathrm{Tr}\big(e^{-\beta H}\big) = \sum_{i=0}^\infty e^{-\beta E_i},
\end{equation}
Because this sum is strictly finite for any inverse temperature $\beta > 0$, the operator $e^{-\beta H}$ rigorously satisfies the definition of a trace-class operator. This absolute convergence provides the foundational mathematical protection, ensuring that the topological index (the Chern character) can be safely evaluated over the infinite-dimensional state space without encountering triviality or divergence \cite{CA1994, SB2005}.

This trace-class rigidity provides the ultimate mathematical justification for the finite-dimensional topological filtration utilized in Section 2. Instead of attempting to define a geometric index directly over the monolithic infinite-dimensional space $\mathcal{H}$, we rigorously decompose the system into the direct sum of its finite-dimensional eigenspaces, $\mathcal{H} = \bigoplus_i \mathcal{H}_i$. Because each eigenspace $\mathcal{H}_n$ associated with an energy level is strictly finite-dimensional, the local vector bundles restricted to these sub-spaces are completely immune to Kuiper's topological triviality.

As demonstrated in the main text, projecting the system onto the lowest $n+1$ spectral modes effectively restricts our geometric evaluation to the projective space $\mathbb{CP}^n$. While the algebraic derivation of the truncated partition function $Z_n(M)$ is straightforward, the critical mathematical subtlety lies in the thermodynamic limit $n \to \infty$. 

For a generic infinite-dimensional bundle, taking this limit would inevitably encounter topological collapse. However, for confined systems, this transition is mathematically shielded precisely by the trace-class norm established above. Because the contributions of eigenspaces of the higher energy level are exponentially suppressed by the statistical weight $e^{-\beta E_n}$, the infinite sum of these finite-dimensional topological contributions absolutely converges. Therefore, the process of directly taking the limit in Section 2 is legal.

\section{Equivariant characteristic classes and algebraic identities}
In this appendix, similar to the exact evaluation of spin path integrals \cite{Ercolessi 1995}, we directly employ local affine coordinates to compute the equivariant characteristic classes on $\mathbb{CP}^n$. The phase space $\mathbb{CP}^n$ is parameterized by the homogeneous coordinates $[z_0 : z_1 : \dots : z_n]$. The thermal evolution operator $e^{-\beta H}$ induces a natural $U(1)$ action on these coordinates:
\begin{equation}z_k 
\mapsto e^{-\beta E_k} z_k, \quad \text{for } k = 0, 1, \dots, n.
\end{equation}
The fixed points of this action correspond exactly to the pure energy eigenstates:
\begin{equation}
p_i = [0 : \dots : 1 : \dots : 0], \quad \text{where only the } i \text{-th coordinate is non-zero.}
\end{equation}
In the affine neighborhood of the fixed point $p_i$ (where $z_i \neq 0$), the $n$ local coordinates spanning the tangent space $T_{\mathbb{CP}^n}(p_i)$ are given by $w_j = z_j / z_i$ for $j \neq i$. Under the $U(1)$ action, these local coordinates transform as:
\begin{equation}
w_j = \frac{z_j}{z_i} \mapsto \frac{e^{-\beta E_j} z_j}{e^{-\beta E_i} z_i} = e^{-\beta(E_j - E_i)} w_j.
\end{equation}
Thus, the equivariant weights (or Chern roots) $\lambda_{ji}$ at $p_i$ are immediately identified as:
\begin{equation}
\lambda_{ji} = -\beta(E_j - E_i), \quad \text{for } j \neq i.
\end{equation}
According to the standard definitions of equivariant characteristic classes, the equivariant Euler class $e_{eq}$ is the product of all weights \cite{atiyah1984moment, tu2020introductory, BNGE VM 1992}, and the equivariant Todd class $\mathrm{Td}_{eq}$ is constructed from the generating function $x/(1-e^{-x})$ \cite{tu2020introductory, BNGE VM 1992}. Evaluating these at the fixed point $p_i$ yields:\\
\begin{itemize}
\item Equivariant Euler class \cite{atiyah1984moment, tu2020introductory, BNGE VM 1992}:
\begin{equation} \label{app_euler}
e_{eq}(T_{\mathbb{CP}^n}(p_i)) = \prod_{j \neq i} \lambda_{ji} = \prod_{j \neq i} \big( -\beta(E_j - E_i) \big).
\end{equation}
\item Equivariant Todd class \cite{BNGE VM 1992}:
\begin{equation} \label{app_todd}
\mathrm{Td}_{eq}(T_{\mathbb{CP}^n}(p_i)) = \prod_{j \neq i} \frac{\lambda_{ji}}{1 - e^{-\lambda_{ji}}} = \prod_{j \neq i} \frac{-\beta(E_j - E_i)}{1 - e^{\beta(E_j - E_i)}}.
\end{equation}
\end{itemize}
Substituting these local classes into the denominator and numerator of the Atiyah-Bott localization formula directly reproduces the rational functions in the main text.

In Section 2, the application of the Atiyah-Bott localization theorem to the phase space $\mathbb{CP}^n$ yields a highly nontrivial localized sum over the fixed points:
\begin{equation} \label{app_sum}
S_n = \sum_{i=0}^n \frac{e^{-\beta E_i}}{\prod_{j \neq i} \left( 1 - e^{-\beta(E_j-E_i)} \right)}.
\end{equation}
To prove that this geometric sum identically collapses into the discrete partition function, we introduce the variable substitution $x_i = e^{-\beta E_i}$. The denominator of the $i$-th term can be algebraically rewritten as:
\begin{equation}
\prod_{j \neq i} \left( 1 - \frac{x_j}{x_i} \right) = \frac{\prod_{j \neq i} (x_i - x_j)}{x_i^n}.
\end{equation}
Note that the product contains exactly $n$ terms since $j$ runs from $0$ to $n$ with $j \neq i$. Substituting this back into the sum, Equation \eqref{app_sum} transforms into:
\begin{equation}
S_n= \sum_{i=0}^n \frac{x_i^{n+1}}{\prod_{j \neq i} (x_i - x_j)}.
\end{equation}
We provide a rigorous and elegant proof of this identity using Cauchy's residue theorem. Consider the following meromorphic function in the complex plane $\mathbb{C}$:
\begin{equation}
f(z) = \frac{z^{n+1}}{\prod_{j=0}^n (z-x_j)}.
\end{equation}
Let $C_R$ be a circular contour centered at the origin with a sufficiently large radius $R$ such that all $n+1$ poles $x_0, x_1, \dots, x_n$ are strictly enclosed within $C_R$. According to the residue theorem, the contour integral of $f(z)$ equals the sum of its residues at the enclosed poles:
\begin{equation} \label{residue_theorem}
\frac{1}{2\pi i} \oint_{C_R} f(z) dz = \sum_{i=0}^n \mathrm{Res}(f, x_i).
\end{equation}
Since each $x_i$ is a simple pole, its residue is directly evaluated by omitting the corresponding $(z - x_i)$ factor in the denominator:
\begin{equation}
\mathrm{Res}(f, x_i) = \lim_{z \to x_i} (z - x_i) f(z) = \frac{x_i^{n+1}}{\prod_{j \neq i} (x_i - x_j)}.
\end{equation}
Thus, the right-hand side of Equation \eqref{residue_theorem} is exactly our localized sum $S_N$.To evaluate the contour integral on the left-hand side, we expand $f(z)$ as a Laurent series for large $|z|$ (as $R \to \infty$). The denominator behaves as an $(n+1)$-th degree polynomial:
\begin{equation} \label{asymptotic form}
\prod_{j=0}^n (z - x_j) = z^{n+1} - \left( \sum_{j=0}^n x_j \right) z^n + \mathcal{O}(z^{N-1}).
\end{equation}
Substituting this asymptotic form \eqref{asymptotic form} into $f(z)$, we obtain:
\begin{equation}
f(z) = \frac{z^{n+1}}{z^{n+1} - \left( \sum_{j=0}^n x_j \right) z^n + \mathcal{O}(z^{n-1})} = \frac{1}{1 - \frac{\sum x_j}{z} + \mathcal{O}(z^{-2})} = 1 + \frac{\sum_{j=0}^n x_j}{z} + \mathcal{O}\left(\frac{1}{z^2}\right).
\end{equation}
When integrated along the large contour $C_R$, the constant term and all $\mathcal{O}(z^{-2})$ terms vanish. The only non-zero contribution arises entirely from the $1/z$ term:
\begin{equation}
\frac{1}{2\pi i} \oint_{C_R} f(z) dz = \frac{1}{2\pi i} \oint_{C_R} \frac{\sum_{j=0}^n x_j}{z} dz = \sum_{j=0}^n x_j.
\end{equation}
Equating the results from the interior residues and the large contour integral, we immediately obtain the exact algebraic identity:
\begin{equation} \label{Residue equation}
\sum_{i=0}^n \frac{x_i^{n+1}}{\prod_{j \neq i} (x_i - x_j)} = \sum_{i=0}^n x_i.
\end{equation}
Restoring the physical variables $x_i = e^{-\beta E_i}$, the geometric sum perfectly collapses into the exact discrete partition function:
\begin{equation}
S_n = \sum_{i=0}^n e^{-\beta E_i}.
\end{equation}
This completes the algebraic proof that the localized continuous geometric integration rigorously reproduces the discrete thermal sum.

\acknowledgments
This work is also supported by NSFC under Grants No. 12275146, the National Key R$\&$D Program of China (2021YFC2203100), the Dushi Program and the Shuimu Fellowship of Tsinghua University. Thanks for helpful discussions with Antonio Padilla, Paul Saffin, Sayyed Rassouli, Jie He, Tillmann Kleiner and Ruizhi Huang and their insightful suggestions. For the purpose of open access, the authors have applied a CC BY public copyright licence to any Author Accepted Manuscript version arising.

\textbf{Conflict of Interest}
\textbf{(The authors declare that there is no conflict of interest.)}

\textbf{Data Availability Statement}
No datasets were generated or analysed during the current study.





\end{document}